# Linear Modeling of the Glass Transition Temperature of the system $SiO_2$-CaO-$Na_2O$

Patrick dos Anjos[1]*
Lucas de Almeira Quaresma*
Marcelo Lucas Pereira Machado*

*Federal Institute of Espírito Santo (IFES), Vitória, ES, Brazil*

## Abstract

This work aimed to mathematically model the glass transition temperature ($T_g$), one of the most important parameters regarding the behavior of slag, being responsible for the sudden change in thermomechanical properties of non-crystalline materials, by the chemical composition of the $SiO_2$-CaO-$Na_2O$ system, widely applicable in the production of glasses and constituent of iron, magnesium and aluminum metallurgy slags. The SciGlass database was used to provide data for mathematical modeling through the Python programming language, using the method of least squares. A new equation was established, called *P Model*, and it presented a lower mean absolute error and lower standard deviation of absolute errors in relation to 3 equations in the literature. The raised equation provides significant results in the mathematical modeling of $T_g$ by the chemical system $SiO_2$-CaO-$Na_2O$, being valid for the limits of the data used in the mathematical modeling.

## 1 Introduction

The glass transition temperature known as $T_g$ is one of the most important parameters in relation to the crystalline behavior of slag, and can be directly obtained through the analysis of the differential exploratory calorimetry (DSC) technique [1]. $T_g$ denotes an abrupt change in the thermomechanical properties of non-crystalline materials, marking the temperature, or its range, where below it the atoms of a supercooled liquid are temporarily frozen, without crystallizing, after cooling [2]. Viscosity is directly related to $T_g$ [3] (Equation 1).

$$\log_{10}(\eta(T, \eta_\infty, T_g, m)) = \log_{10}(\eta_\infty) + T_g/T\,(12 - \log_{10}(\eta_\infty))\,\exp((T_g/T - 1)\,(m/(12 - \log_{10}(\eta_\infty)) - 1) \quad (1)$$

The parameter m known as the fragility index is a relationship between viscosity (η) and $T_g$, as shown in Equation 2.

1 E-mail: patrick.dosanjos@outlook.com

$$m = \partial \log_{10}(\eta(T))/\partial(T_g/T) \,\Big|\, T=T_g \quad (2)$$

The $SiO_2$-$CaO$-$Na_2O$ system is one of the chemical systems used for application in glasses and containers [4], being components in blast furnace slag [5], in the slag of the Pidgeon process [6] and in the red mud resulting from the leaching of aluminum ores [7]. Slag applied in continuous casting may have glass phases [8], and the study of the glass transition temperature can be valuable.

Mathematical modeling of $T_g$ can be done through chemical composition [9-10]. Mills et al. (2016) described a linear equation that correlates $T_g$ in the $SiO_2$-$CaO$-$Al_2O_3$-$Na_2O$-$K_2O$-$MgO$-$CaF_2$-$MnO$-$FeO$ system, modeled through $T_g$ data from 23 different mould slags, resulting in Equation 3, which has uncertainty of approximately 20K (X = mole fraction).

$$T_g(K) = 906 - 330.5X\mathit{SiO_2} + 190X\mathit{CaO} + 440X\mathit{Al_2O_3} - 440X(Na_2O+K_2O) - 11X\mathit{MgO} + 154X\mathit{CaF_2} - 309X\mathit{MnO} - 1391X\mathit{FeO} \quad (3)$$

Andersson (1992) established the glass transition temperature in a $SiO_2$-$Na_2O$-$CaO$-$P_2O_5$-$Al_2O_3$-$B_2O_3$ system, which can be calculated by Equation 4 or Equation 5. These equations were modeled through 16 different glasses, providing mean absolute error of 6.7°C for Equation 4 and 8.2°C for Equation 5 (% = mass percentage).

$$T_g(°C) = 44.3136 + 6.27159\%SiO_2 + 6.90715\%CaO + 6.32101\%P_2O_5 + 6.75157\%Al_2O_3 \quad (4)$$

$$T_g(°C) = 674.483 - 6.26843\%Na_2O \quad (5)$$

This work aimed to perform a linear modeling of the glass transition temperature correlating the chemical composition of the $SiO_2$-$CaO$-$Na_2O$ system to provide an equation with low error and low variability.

## 2 Materials and Methods

For the mathematical modeling of the $SiO_2$-$CaO$-$Na_2O$ system relating to $T_g$, the SciGlass database was used [3]. For data modeling, the treatments of missing values (NaN) were performed, the removal of a sub-database where the sum of the elements $SiO_2$, $CaO$ and $Na_2O$ completed 100%, with subsequent conversion of the chemical species to molar fraction and the conversion of $T_g$ values to Kelvin. The resulting sub-database had the $SiO_2$-$CaO$-$Na_2O$ system relating each chemical system to its $T_g$. The sub-database parameters are seen in Table 1.

Table 1. Parameters of the sub-database used for linear modeling.

| **Parameters** | **$SiO_2$ (%mol)** | **$CaO$ (%mol)** | **$Na_2O$ (%mol)** | **$T_g$ (K)** |
|---|---|---|---|---|
| Data | 512 | 512 | 512 | 512 |
| Mean | 68.13 | 12.23 | 19.64 | 835.40 |
| Standard Deviation | 14.34 | 17.64 | 13.90 | 160.78 |
| Minimum | 31.25 | 0.00 | 0.00 | 579.15 |
| Maximum | 100.00 | 61.00 | 68.75 | 1495.15 |

Isolated chemical species do not have a linear mathematical relationship with $T_g$, as can be seen in Figure 1. Hence a simple linear regression does not have statistical representation.

Mathematical modeling was performed using the Python programming language by applying the least squares method, by minimizing and generalizing Equation 6 (*a*=linear coefficient, *b*=angular coefficients vector, *y*=dependent variable, *x*=independent variables vector).

$$f(a, b) = \Sigma(y - a - bx)^2 \quad (6)$$

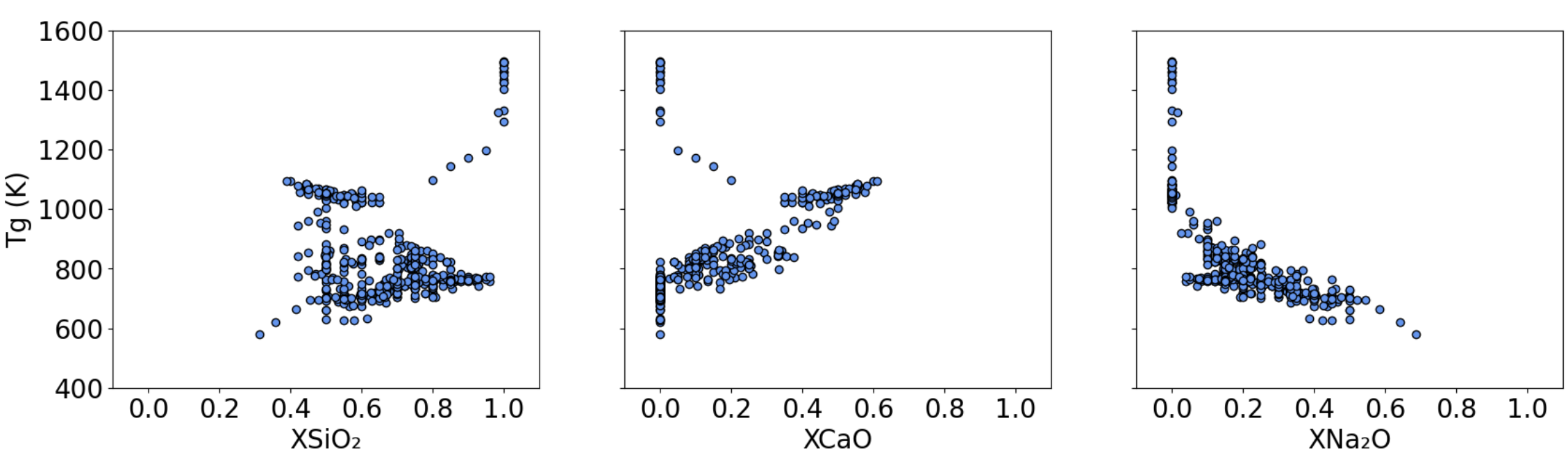

Figure 1. Relationship between $SiO_2$, CaO and $Na_2O$ separately with $T_g$ (X = molar fraction).

The evaluations of the new model were performed by statistical methods of mean absolute error (MAE) (Equation 7) and the standard deviation of the absolute error between the results of the new model in relation to the sub-database (STD) (Equation 8).

$$\text{MAE (K)} = 1/N \, \Sigma \left| y\textit{predicted} - y\textit{true} \right| \quad (7)$$

$$\text{STD (K)} = \sqrt{\text{Var}}( \left| y\textit{predicted} - y\textit{true} \right| ) \quad (8)$$

For mathematical consistency, the evaluation was also performed in the model described by Mills et al. (2016) and by both models developed by Andersson (1992).

## 3 Results and discussion

The result of the multiple regression resulted in the values of a=776.62, b=180.77, c=332.45 and d=-513.22 and the linear equation resulting from the mathematical modeling is denoted by Equation 9, named *P model* (X = molar fraction).

$$T_g\,(K) = 776.62 + 180.77X\mathit{SiO_2} + 332.45X\mathit{CaO} - 513.22X\mathit{Na_2O} \quad (9)$$

The statistical parameters of mean absolute error (MAE) and standard deviation of absolute error (STD) are seen in Table 2.

Table 2. MAE and STD statistical parameters of Equations 3, 4, 5 and 9.

| **Models** | **Mills et al. (2016) (Equation 3)** | **Andersson (1992) (Equation 4)** | **Andersson (1992) (Equation 5)** | **P Model (Equation 9)** |
|---|---|---|---|---|
| MAE (K) | 219.49 | 62.72 | 67.68 | 58.40 |
| STD (K) | 126.59 | 88.09 | 88.67 | 86.35 |

P Model exhibited MAE of 58.40K and STD of 86.35K and the best model among those mentioned was Equation 4 of Andersson (1992) with MAE of 62.72K and STD of 88.09K.

Among the 4 models, the P Model was considered the model with the lowest error and lowest variability. The Andersson (1992) model with the lowest error and lowest variability was Equation 4, being modeled with 16 different glasses, whereas the P Model was built from 520 different chemical composition data of the $SiO_2$-CaO-$Na_2O$ system relating the $T_g$. Therefore, the P Model has a greater statistical representation, since it used a greater amount of data.

With the results predicted in the 2 best models, the P model and Equation 4 described by Andersson (1992), a scatter plot of errors was generated in relation to the sub-database (Figure 2).

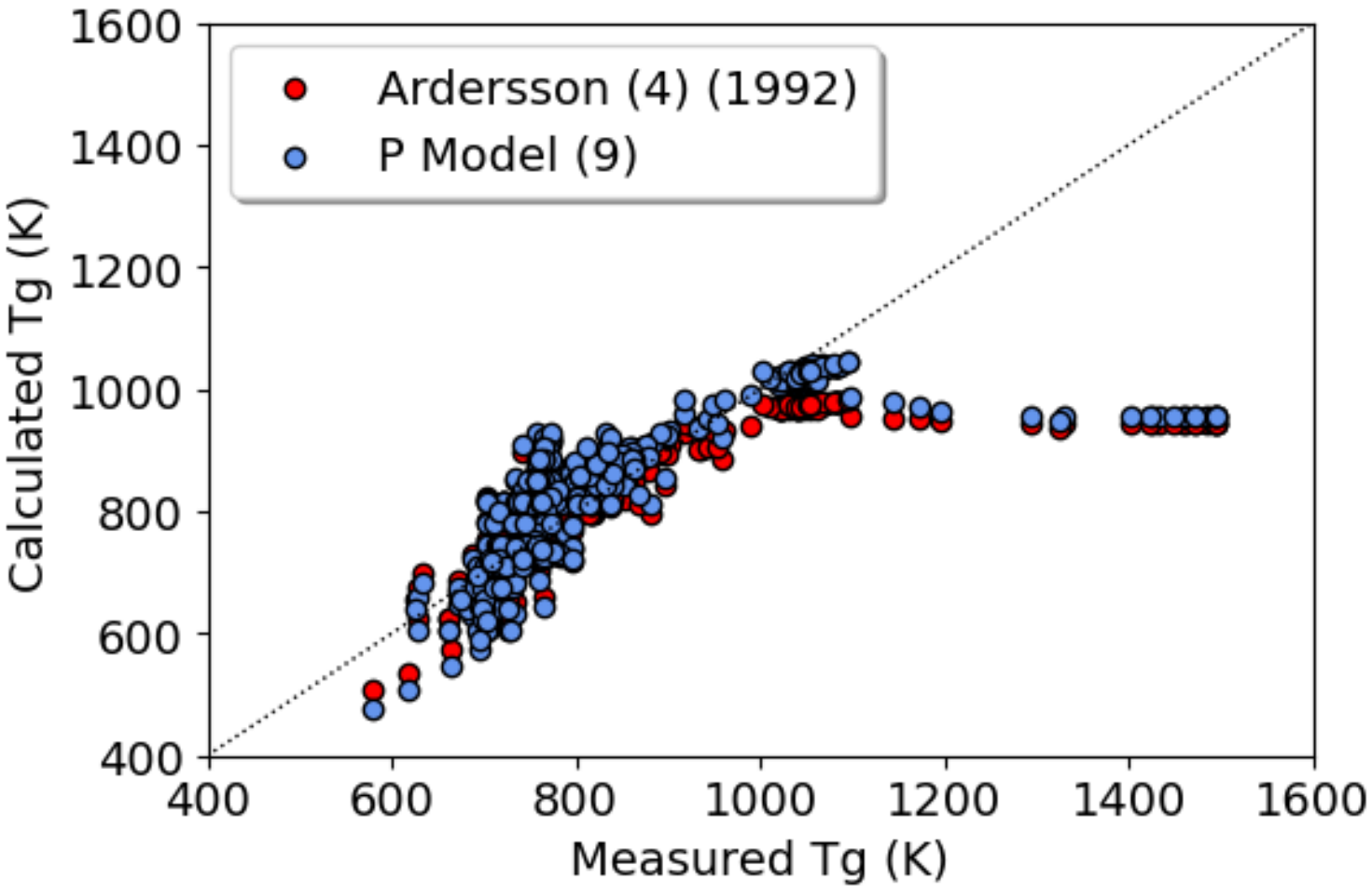

Figure 2. Scatter plot of errors between Equations 4 and 9 in relation to the sub-database.

The sensitivity of the system regarding the constants related to the chemical species remained the same, with the molar fraction of $SiO_2$ and CaO increasing $T_g$ and $Na_2O$ decreasing in the $SiO_2$-$Na_2O$-CaO-$P_2O_5$-$Al_2O_3$-$B_2O_3$ and $SiO_2$-CaO-$Na_2O$ systems. The deviation values of the P Model are shown in Figure 3.

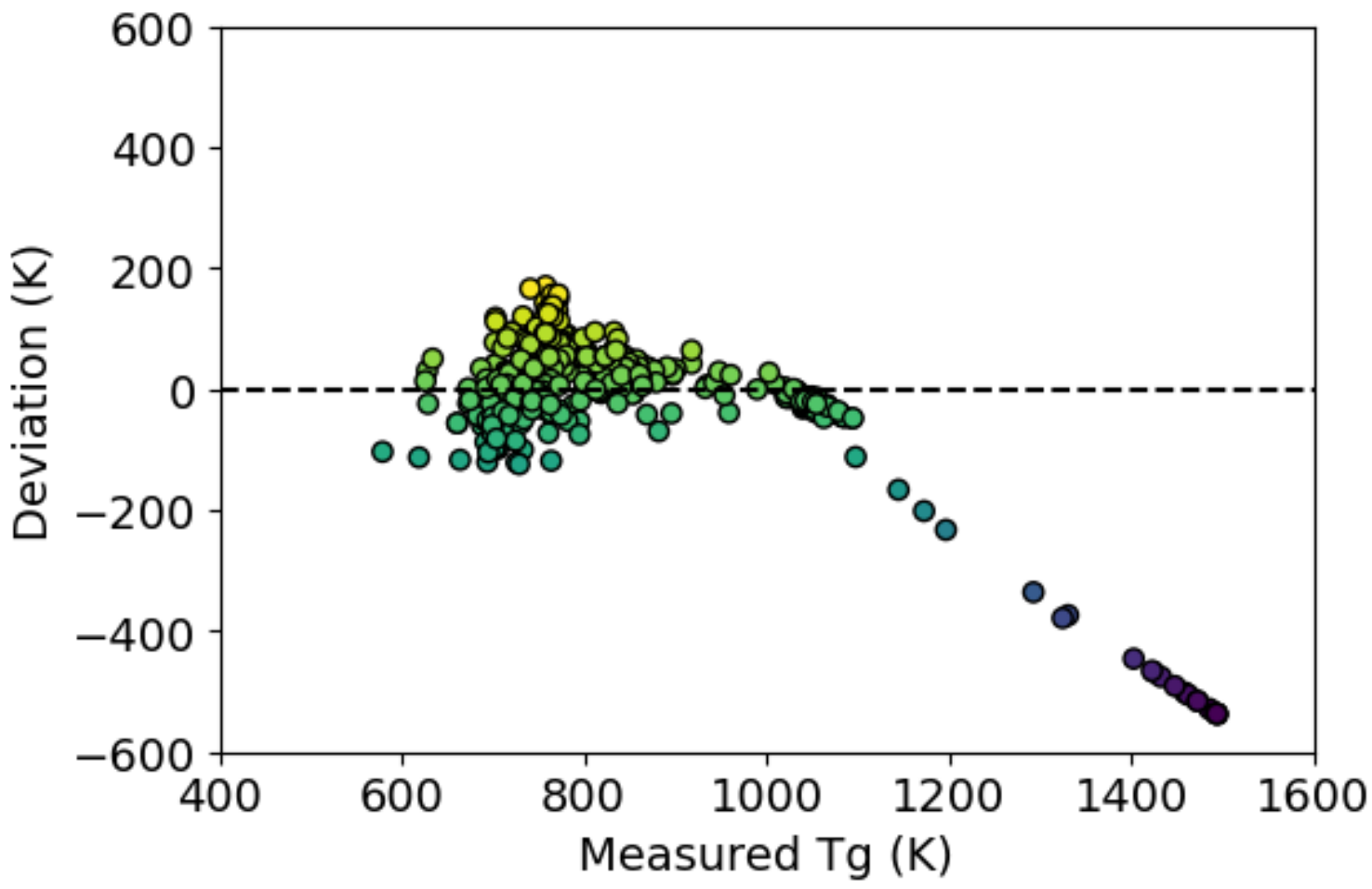

Figure 3. P Model deviation values.

By analysis, it can be seen that there is an increase in the deviation values when the $T_g$ is greater than 1000K, which denotes that the $T_g$ has more predictors than just the chemical composition.

## 4 Conclusion

Glass transition temperature ($T_g$) values were modeled using the SciGlass database, the predictor variables being the chemical composition of the $SiO_2$-CaO-$Na_2O$ system. Using a linear modeling, the equation $T_g$ (K) = 776.62 + 180.77X*$SiO_2$* + 332.45X*CaO* – 513.22X*$Na_2O$* called *P Model* was obtained relating the chemical composition of the system in molar fraction. The P Model was statistically evaluated and resulted in a mean absolute error (MAE) of 58.40K and standard deviation of absolute error (STD) of 86.35K and the best model in literature, the model of Andersson (1992) represented by Equation 4, resulted in a MAE of 62.72K and STD of 88.09K, demonstrating that the P Model has less error and less variability in the 4 equations described. With a sub-database composed of 520 different data P Model was modeled, in contrast to the best model in the literature, Andersson (1992) represented by Equation 4, which was modeled with 16 different data, denoting that the P Model has a greater statistical representation. analyzing the results, the values of $T_g$ > 1000K have deviation values greater than the $T_g$ < 1000K, demonstrating a threshold of divergence in the P Model, indicating that above this threshold the $T_g$ has more predictive elements than just the chemical composition.